\documentclass[10pt,a4paper]{article}             

\usepackage{amsmath,amsthm,amssymb,mathrsfs,bm,bbm,enumerate,physics}
\usepackage{tikz}
\usetikzlibrary{intersections,calc,arrows.meta}
\usepackage{float,placeins,adjustbox}
\usepackage[hyphens]{url}
\usepackage{hyperref}
\usepackage{cleveref}

\usepackage{comment}
\usepackage{caption}
\newcommand\myshade{75}
\colorlet{mylinkcolor}{red}
\colorlet{mycitecolor}{green}
\colorlet{myurlcolor}{blue}
\hypersetup{colorlinks=true, anchorcolor=cyan, 
linkcolor=mylinkcolor!\myshade!black, 
citecolor=mycitecolor!\myshade!black,
urlcolor=myurlcolor!\myshade!black } 

\newcommand{\beq}{\begin{equation}}
\newcommand{\eeq}{\end{equation}}

\newcommand{\ra}{\rightarrow}
\newcommand{\lra}{\leftrightarrow}
\newcommand{\qqi}{q-q^{-1}}
\newcommand{\eps}{\epsilon}
\newcommand{\scz}{\widehat{\mkern-4mu CZ}}

\newcommand{\Op}[3]{ \hat{#1}_{#2}^{({#3})}}

\theoremstyle{definition}
\newtheorem*{rmk*}{$\ddagger$  Remark}

\theoremstyle{remark}
\newtheorem*{exm*}{$\spadesuit$ Example}

\makeatletter
\long\def\@makefntext#1{\parindent 1em\noindent
\@hangfrom{\hbox to 1.8em{\hss$^{\@thefnmark}$}}#1}
\makeatother
\begin{document}
\topmargin 0pt
\oddsidemargin 0mm
\renewcommand{\thefootnote}{\fnsymbol{footnote}}

\vspace*{0.5cm}

\begin{center}
{\Large Curtright-Zachos Supersymmetric Deformations of the Virasoro algebra in Quantum Superspace and Bloch Electron Systems}
\vspace{1.5cm}

{\large Haru-Tada Sato${\,}^{a,b}$
\footnote{\,\,\, Corresponding author.  E-mail address: satoh@isfactory.co.jp}
}\\

{${}^{a}$\em Department of Physics, Graduate School of Science, \\Osaka Metropolitan University \\
Nakamozu Campus, Sakai, Osaka 599-8531, Japan}\\

{${}^{b}$\em Department of Data Science, i's Factory Corporation, Ltd.\\
     Kanda-nishiki-cho 2-7-6, Tokyo 101-0054, Japan}\\
%
\end{center}

\vspace{0.1cm}

\abstract{
We introduce supersymmetric extensions of the Hom-Lie deformation of the Virasoro algebra, as realized in the GL(1,1) quantum superspace, for Bloch electron systems under Zeeman effects. The construction is achieved by defining generators through magnetic translations and spin matrix bases, specifically for the 
$N=1$ and $N=2$ supersymmetric deformed algebras. This approach reveals a structural parallel between the deformed algebra in quantum superspace and its manifestation in Bloch electron systems.
}
\vspace{0.5cm}

\begin{description}
\item[Keywords:] quantum superspace, $q$-deformation, Virasoro algebra, Hom-Lie algebra, Zeeman effect, tight binding model 
\item[MSC:] 17B61, 17B68, 81R50, 81R60
\end{description}

%
%

%
\newpage
\setcounter{page}{1}
\setcounter{footnote}{0}
\setcounter{equation}{0}
\setcounter{secnumdepth}{4}
\renewcommand{\thepage}{\arabic{page}}
\renewcommand{\thefootnote}{\arabic{footnote})}

The study of noncommutative geometry has emerged as a fundamental framework in physics, particularly in systems under strong external fields such as magnetic fields. This field encompasses various phenomena including noncommutative field theory~\cite{NCFT,SW}, quantum Hall states~\cite{latestQH}, AdS/CFT correspondence, and black hole physics~\cite{AdS,Edge,BHS,Strom}. The mathematical tools employed to describe noncommutative spaces primarily fall into two categories: Moyal deformations, which introduce noncommutativity in flat or phase spaces~\cite{Moyal}, and quantum spaces associated with quantum groups~\cite{ma,tak,maj}. While Moyal deformation preserves classical symmetry and group structures with providing a framework for field theories and quantum mechanical actions through the Moyal product, quantum group approaches offer distinct mathematical frameworks with different types of noncommutativity.

Recent developments have revealed intriguing connections between quantum space and the Curtright-Zachos (CZ) deformation of the Virasoro algebra~\cite{CZ} in physical systems~\cite{AS2}, particularly in relation to Moyal deformation~\cite{AS3}. When magnetic translation (MT) operators are expressed in the angular momentum representation on a cylinder, the linear-combined operators $L_n$ manifest as one-dimensional $q$-difference operators. These operators exhibit commutation relations analogous to those of differential operators on quantum planes associated with quantum groups, establishing the CZ algebra as a bridge between Moyal deformation and the noncommutativity of the quantum plane~\cite{AS2,AS3}.

The CZ algebra has spawned numerous theoretical developments, including central extensions and operator product formula (OPE)~\cite{AS,Poisson}, supersymmetric extensions~\cite{MZ,superCZ,superCZ2,super3,HHT}, multi-parameter deformations~\cite{pqCZ,pqCZ2}, and recently deformation of open string field theory~\cite{qstr} (see also Section 1.3 in \cite{AS2} for further references).

The CZ algebra was originally proposed by Curtright and Zachos as a $q$-deformation of the Virasoro algebra~\cite{CZ}:
\beq
[L_n,L_m]_\ast=(L_nL_m)_\ast-(L_mL_n)_\ast=[n-m]L_{n+m}\,, \label{CZ}
\eeq
where $(L_nL_m)_\ast=q^{m-n}L_nL_m$ and the $q$-bracket symbol $[A]$ is defined as
\beq
[A]=\frac{q^A -q^{-A}}{\qqi}\,,\quad \mbox{where}\quad q=e^{i\hbar\omega} \,,
\label{qbraA}
\eeq
This is mathematically interpreted as the Hom-Lie algebras~\cite{Hom1,Hom2,Hom3}. While attempts have been made to construct CZ algebra using quantum planes covariant under quantum groups~\cite{superCZ,superCZ2,HHT,HHTZ}, particularly in the context of supersymmetry, these approaches have primarily focused on the quantum group structure itself, leaving the physical manifestation of these structures and their role in CZ algebra emergence largely unexplored.

Our work addresses this gap by focusing on physical systems that naturally exhibit CZ algebra and investigating their supersymmetric extensions. Building upon existing research on super CZ algebras based on quantum superspaces (QSS) covariant under $OSp_q(1,2)$ and $GL_q(1,1)$~\cite{superCZ,superCZ2}, we demonstrate how these structures manifest in two-dimensional electron systems under magnetic fields. This approach provides a concrete physical realization of the abstract mathematical frameworks previously developed in the quantum group context.

Specifically, we investigate supersymmetric extensions of the Hom-Lie deformation of the Virasoro algebra within the $GL_q(1,1)$ quantum superspace framework, particularly as it applies to Bloch electron systems under Zeeman effects. Our construction utilizes generators defined through magnetic translations and spin matrix bases, focusing on $N=1$ and $N=2$ supersymmetric deformed algebras. This approach reveals fundamental connections between the abstract mathematical structure in quantum superspace and their physical manifestation in Bloch electron systems.

\section{Magnetic translation (MT) and $CZ$ algebras}
\label{sec:CZ}
\indent

The CZ algebra \eqref{CZ} represents a Hom-Lie deformation of the Virasoro algebra, 
which can be realized through magnetic translation (MT) operators~\cite{AS2}. 
In addition to the $CZ^+$ algebra defined in \eqref{CZ}, there are two 
other variations: $CZ^-$ and $CZ^\ast$. 
The algebras $CZ^\pm$ exhibit symmetry under the interchange $q\lra q^{-1}$, 
while the $CZ^\ast$ algebra serves an extended algebra encompassing both $CZ^\pm$. 
The MT operators satisfy the Moyal commutation relation
\beq
[T_n^{(k)}\,, T_m^{(l)}]=[\frac{nl-mk}{2}]T_{n+m}^{(k+l)}  \,,  \label{eq:FFZ}
\eeq
which derives from two fundamental relations: the MT exchange relation
\beq
  \hat{T}_n^{(k)}\hat{T}_m^{(l)}= q^{nl-mk}  \hat{T}_m^{(l)}\hat{T}_n^{(k)} \,,
\eeq
and the MT fusion relation
\beq
\hat{T}_n^{(k)}\hat{T}_m^{(l)}= \frac{1}{\qqi} q^{\frac{nl-mk}{2}}  \hat{T}_{n+m}^{(k+l)}\,.  \label{Trans}
\eeq
The MT operators $\Op{T}{n}{k}$ can be expressed as differential operators, structured as $\Op{T}{n}{0}$ multiplied by $\hat{S}_0=q^{-2z\partial}$ from left (with a free parameter $\Delta$):
\beq
\Op{T}{n}{k}=q^{k(\frac{n}{2}-\Delta)}\hat{S}_0^{\frac{k}{2}}\Op{T}{n}{0}
=\frac{z^n}{\qqi} q^{-k(z\partial+\frac{n}{2}+\Delta)}\,.  \label{Tnk20}
\eeq

The $CZ^\pm$ generators $\hat{L}_n^\pm$ are constructed from MT operators 
with weight pair $(0,\pm2)$~\cite{AS2}:
\beq
\hat{L}_n^\pm = \mp\Op{T}{n}{0} \pm q^{\pm(n+2\Delta)} \Op{T}{n}{\pm2}\,,  \label{Lpm1} 
\eeq
which can be expressed in $q$-difference form as
\beq
\hat{L}_n^\pm = \mp z^{n}\frac{1-q^{\mp2z\partial}} {\qqi} 
=:-z^{n+1}\partial_q^\pm  \,.  \label{Ln_diff}
\eeq
The MT and $CZ^\pm$ algebras satisfy the following $\ast$-bracket commutation relations:
\begin{align}
&    [\Op{T}{n}{k},\Op{T}{m}{l}]_\ast =0\,,                 \label{*1}     \\
& [\hat{L}_n^\pm, \hat{L}_m^\pm]_\ast = [n-m]\hat{L}_{n+m}^\pm\,,   \label{*2}   \\
&  [\hat{L}_n^\pm,\Op{T}{m}{l}]_\ast = -[m]\Op{T}{n+m}{l}\,.  \label{*3}  
\end{align}
The $CZ^\ast$ algebra, describing interactions between $CZ^\eps$ generators 
($\eps=\pm$), takes the form: 
\beq
 [L_n^\eps,L_m^\eta]_\ast =q^{\eta m}[n]L_{n+m}^\eps - q^{\eps n}[m]L_{n+m}^\eta\,,  \label{CZCZ}
\eeq
where the $\ast$-bracket commutator
\beq
[X_n^{(k)},X_m^{(l)}]_\ast=(X_n^{(k)}X_m^{(l)})_\ast-(X_m^{(l)}X_n^{(k)})_\ast \label{*4}
\eeq
is defined for every element 
$X_n^{\eps(k)} \in \mathscr{M}_T\{\hat{L}_n^\eps,T_n^{(k)}\}$ of weight $k$ by 
\beq
(X_n^{\eps(k)}X_m^{\eta(l)})_\ast = q^{-x(\eps,\eta)} X_n^{\eps(k)} X_m^{\eta(l)}\,,
\quad\,\,  x(\eps,\eta)=\frac{\eta nl-\eps mk}{2}  \label{X*X}
\eeq
where $k$ and $l$ in $x(\eps,\eta)$ should be set to 2 for $\hat{L}_n^\eps$ and $k$ 
for $\hat{T}_n^{(k)}$. The signature symbol $\eps$ denotes $\pm$ for $\hat{L}_n^\pm$ 
and $\eps=+$ for $\hat{T}_n^{(k)}$. Furthermore, $\hat{S}_0$ acts as a central 
element when we condider the weight of $\hat{S}_0^k$ as $2k$ in \eqref{X*X} for $k\not=0$~\cite{AS} 
\beq
[\hat{S}_0^k, \hat{L}_n]_\ast =0\,,\quad [\hat{S}_0^k, \hat{T}_m^{(l)}]_\ast =0\,.
\eeq

\section{Virasoro superalgebra in Bloch electron system}\label{sec:SV}
\indent

Since the existence of Grassmann bases is essential for constructing superalgebras, 
we consider a one-electron spin system in a static magnetic field with spin-magnetic interaction (Zeeman term) as a typical quantum mechanical system with built-in Grassmann bases.
\beq
H=\frac{1}{2m}(\bm{\sigma}\cdot\bm{\pi})^2=\frac{1}{2m}\bm{\pi}^2
+\frac{1}{2}g \mu_B \bm{\sigma}\cdot \bm{B}
\eeq
Here, $\sigma_i$, $\mu_B$, and $g$ are the Pauli matrices, Bohr magneton, and $g$-factor, respectively, where
\beq
\mu_B=\frac{e\hbar}{2mc}\,,\quad g=2(1+\frac{\alpha}{2\pi}+O(\alpha^2))\,,\quad
\alpha=\frac{e^2}{\hbar c}
\eeq
($\alpha$ is the fine structure constant $\approx1/137$). When neglecting relativistic effects, $g=2$, and since this is not essential for the following discussion, we set $g=2$.
Furthermore, taking $\bm{B}=(0,0,B)$, we write
\beq
H=H_0 + \delta H\,,\quad H_0=\frac{1}{2m}\bm{\pi}^2 \,,\quad 
\delta H=\mu_B B\sigma_z\,.
\eeq
When there exists a base operator $\mathcal{O}_\beta$ (for example $\hat{T}_n^{(k)}$) that commutes with $H_0$, the following construction forms bases that commute with $H$ (i.e., commute with $\sigma_z$):
\beq
\mathcal{O}_\beta\otimes 1\,,\quad \mathcal{O}_\beta\otimes\sigma_z\,,\quad
\mathcal{O}_\beta\otimes\sigma_1\sigma_2\,,\quad \mathcal{O}_\beta\otimes\sigma_2\sigma_1\,.
\eeq
We call these the magnetic spin base (MSB) in this papar. Here,
\beq
\sigma_1\sigma_2=\begin{pmatrix}
0 & 0  \\
0 & 1  \\
\end{pmatrix} \,,\quad\quad
\sigma_2\sigma_1=\begin{pmatrix}
1 & 0  \\
0 & 0  \\
\end{pmatrix} 
\eeq
and $\sigma_1,\sigma_2$ constitute Grassmann bases that anticommute with $\delta H$ (i.e., anticommute with $\sigma_z$):
\beq
\{\sigma_1,\sigma_2\}=1\,,\quad \sigma_1^2=\sigma_2^2=0\,,
\eeq
where
\beq
\sigma_1=\sigma_x-i\sigma_y=\begin{pmatrix}
0 & 0  \\
1 & 0  \\
\end{pmatrix} \,,\quad\quad
\sigma_2=\sigma_x+i\sigma_y=\begin{pmatrix}
0 & 1  \\
0 & 0  \\
\end{pmatrix}\,.
\eeq
These Grassmann bases can be associated with Grassmann variables and their derivatives in superspace $(x,\theta)$. We refer to this as the SSM correspondence ( acronym from superspace and spin matrix):
\beq
\sigma_1\leftrightarrow \theta\,,\quad 
\sigma_2\leftrightarrow \partial_\theta\,.  \label{grass}
\eeq

Before considering the super CZ algebra in QSS, we first review the process of deriving the Virasoro superalgebra in MSB space by applying the SSM correspondence \eqref{grass} to the conventional Virasoro superalgebra on ordinary superspace (SS) as our starting point. Here, we deal with a situation where the magnetic field is very weak, allowing us to take $q=1$, and the Virasoro superalgebra is realized in a state where only supersymmetry remains. First, we prepare the Virasoro operator $V_n$ and $U(1)$ operator $F_n$ as bosonic fundamental operators:
\beq
V_n=-x^{n+1}\partial_x\,,\quad F_n=x^n\,.
\eeq
Let $J_n$ denote the super current obtained by applying the operator $F_n$ (which represents a component of the dilatation on bosonic space) to the superspace. Then, involving the super Virasoro operator $L_n$ and the supercharge $G_r$, we have
\begin{align}
&L_n=V_n-\frac{n+1}{2}J_n\,,  \label{EYform1} \\
&J_n=F_n\theta\partial_\theta\,,\quad G_r=x^{r+\frac{1}{2}}(\partial_\theta-\theta\partial_x)\,,
\end{align}
and the Virasoro super algebra ($N=1$) is realized:
\begin{align}
&[L_n,L_m]=(n-m)L_{n+m}\,,\quad \{G_r,G_s\}=2L_{r+s}\,, \label{SVfrom}  \\
&[L_n,G_r]=(\frac{n}{2}-r)G_{n+r}\,, \\
&[L_n,J_m]=-mJ_{n+m}\,,\quad [J_n,J_m]=0\,.
\end{align}
The decomposition into $N=2$ super Virasoro algebra is given by
\beq
G_r=G_r^+ + G_r^-\,,\quad G_r^-=x^{r+\frac{1}{2}}\partial_\theta\,,\quad 
G_r^+=-x^{r+\frac{1}{2}} \theta \partial_x\,, 
\eeq
satisfying the following relations:
\begin{align}
&\{G_r^\pm,G_s^\pm\}=0\,,\quad \{G_r^+,G_s^-\}=L_{r+s}+\frac{1}{2}(r-s)J_{r+s}\,, \\
&[L_n,G_r^\pm]=(\frac{n}{2}-r)G_{n+r}^\pm\,, \quad[J_n,G_r^\pm]=\pm G_{n+r}^\pm\,.
\label{SVto}
\end{align}
By applying the SSM correspondence \eqref{grass} to $G_r,J_n,L_n$, we can obtain their MSB representation
\begin{align}
&\mathcal{G}_r=V_{r-\frac{1}{2}}\otimes\sigma_1 + F_{r+\frac{1}{2}}\otimes\sigma_2\,,
\label{Gform2} \\
&\mathcal{J}_n=F_n\otimes\sigma_1\sigma_2\,, \label{Jform2} \\
&\mathcal{L}_n=V_n\otimes1 -\frac{n+1}{2}F_n\otimes\sigma_1\sigma_2\,,
\label{EYform2}
\end{align}
and their explicit matrix representations are as follows:
\beq
\mathcal{G}_r=\mathcal{G}_r^+ + \mathcal{G}_r^-\,,\quad
\mathcal{G}_r^+=\begin{pmatrix}
0 &  0  \\
V_{r-\frac{1}{2}} & 0  \\
\end{pmatrix} \,,\quad\quad
\mathcal{G}_r^-=\begin{pmatrix}
0 &  F_{r+\frac{1}{2}}  \\
0 & 0  \\
\end{pmatrix} \,,
\eeq
\beq
\mathcal{J}_n=\begin{pmatrix}
0 & 0  \\
0 & F_n  \\
\end{pmatrix} \,,\quad\quad
\mathcal{L}_n=\begin{pmatrix}
V_n & 0  \\
0 & L_n  \\
\end{pmatrix} \,.\quad\quad \label{BL2let}
\eeq
These $\mathcal{L}_n,\mathcal{G}_r,\mathcal{G}_r^\pm, \mathcal{J}_n$ satisfy the above Virasoro super algebra ($N=1,2$) \eqref{SVfrom}-\eqref{SVto}.

\section{super CZ algebra in quantum superspace (QSS)}\label{sec:SCZ}
\indent

Now let us consider the case where superspace $(x,\theta)$ is replaced with QSS. 
Here we deal with two types of (super) CZ algebras \cite{superCZ,superCZ2} in quantum space that are $GL_q(1,1)$ conjugate. The reason for considering two types is because there exist two types of Virasoro operators in the diagonal components of the MSB representation \eqref{BL2let}.
First, taking the bosonic CZ operator $B_n$ and U(1) operator $J_n$ as fundamental operators
\beq
B_n=-q^{-1}x^{n+1}\partial_x\,,\quad J_n=x^n\theta\partial_\theta\,, \label{Bn}
\eeq
they satisfy the following relations:
\begin{align}
&[B_n,B_m]_{(m-n)}=[n-m]B_{n+m}\,,\label{BBBFq} \\
&[B_n,J_m]_{(m-n)}=-q^{-n}[m] J_{n+m}\,, \quad
[J_n,J_m]_{(m-n)}=0\,.  \label{FFU1q}
\end{align}
The first CZ operator (type 1) is a composite operator of $B_n$ and $J_n$ as
\beq
L_n=B_n-g_n J_n\,,\quad g_n=a q^{-2n}+b :=g_n^{CZ}\,, \label{CZgen}
\eeq
where $L_n$ satisfies the same commutation relations as $B_n$~\cite{superCZ2}:
\beq
[L_n,L_m]_{(m-n)}=[n-m]L_{n+m}\,,\quad[L_n,J_m]_{(m-n)}=-q^{-n}[m]J_{n+m}\,.
\eeq
The zero mode $L_0$ is related to the bosonic scaling operator $\mu$,
\beq
\mu=\partial_x x-x\partial_x=1+(q-q^{-1})L_0 \,. \label{mubyL0}
\eeq
The first CZ operator corresponds to the counterpart of $V_n$ in \eqref{BL2let}.

For the second (super) CZ operator, we consider a slightly different composition $L'_n$ from \eqref{CZgen}~\cite{superCZ}:
\beq
L'_n=B_n-g'_n J_n\,,\quad g'_n=a' q^{-n}+b'\,. \label{CZTgen}
\eeq
The commutation relations for $L'_n$ receive the following modification (which we denote as $\scz$):
\begin{align}
&[L'_n,L'_m]_{(m-n)}=[n-m]L'_{n+m}+a_{n,m}J_{n+m}\,, \label{CZtwist} \\
&[L'_n,J_m]_{(m-n)}=-q^{-n}[m]J_{n+m}\,,
\end{align}
where
\beq
a_{n,m}=a'q^{-n-m}([m-n]+[n]-[m])=a'c^2q^{-n-m}
[\frac{n-m}{2}][\frac{n}{2}][\frac{m}{2}] \,,  \label{a_nm}
\eeq
\beq
c=\qqi\,.  \label{c}
\eeq
The super $\scz$ algebra in QSS takes the supercharge and $g'_n$ as follows:
\beq
G_r=\lambda^{-\frac{1}{2}}x^{r+\frac{1}{2}}(\partial_\theta-\theta\partial_x)\,, \label{Gtype1}
\eeq
where
\beq
\lambda=1+(q-q^{-1})L'_0\,,
\eeq
and
\beq
g'_n=q^{-\frac{n+1}{2}}[\frac{n+1}{2}]\,.
\eeq
In super $\scz$, the algebra closes in a simple form for the anticommutation relations of $G_r$:
\beq
[L'_n,G_r]_{(r-\frac{n}{2})}=q^{-n}[\frac{n}{2}-r]G_{n+r}\,, \label{LnGr2}
\eeq
\beq
\{G_r,G_s\}_{(\frac{s-r}{2})}=q^{r+s+\frac{5}{2}}(q^{\frac{s-r}{2}}+q^{\frac{r-s}{2}})L'_{r+s}\,.
\label{GrGs2}
\eeq
The super $\scz$ algebra \eqref{LnGr2} and \eqref{GrGs2} can be decomposed into $N=2$ algebra \cite{superCZ}:
\beq
G_r=G_r^++G_r^-\,,    \label{N2Gr}
\eeq
\beq
G_r^-=\lambda^{-\frac{1}{2}}x^{r+\frac{1}{2}}\partial_\theta\,,\quad
G_r^+=-\lambda^{-\frac{1}{2}}x^{r+\frac{1}{2}}\theta\partial_x\,,  \label{N2Grpm}
\eeq
\beq
\{G_r^+,G_s^-\}=q^{r+s+\frac{5}{2}}L'_{r+s}+q^{\frac{r-s+3}{2}}[\frac{r-s}{2}]J_{r+s}\,\,,\quad
\{G_r^\pm,G_s^\pm\}=0\,,  \label{N2GG}
\eeq
\beq
[L'_n,G_r^\pm]_{(r-\frac{n}{2})}=q^{-n}[\frac{n}{2}-r]G_{n+r}^\pm\,, \label{N2LG}
\eeq
\beq
[J_n,G_r^+]_{(\alpha,\beta)}=q^{n+2+\alpha} \lambda G_{n+r}^+ \,, \quad
[J_n,G_r^-]_{(\alpha,\beta)}=-q^{n+2r+1+\beta} \lambda G_{n+r}^-\,. \label{N2JG}
\eeq

We do not yet know how to fix $\alpha$ and $\beta$. As we will see later, these free values will be fixed when we impose some physical requirements, for example, the requirement that they should correspond to the super CZ of the electron spin system.

\section{super CZ algebra in the electron spin system}\label{sec:spin_rep}
\indent

As the first step, we show that the CZ algebra in quantum space (QS), $CZ_{QS}$, can be transformed into the magnetic translation CZ algebra, $CZ_{MT}$, by utilizing the correspondence between the QS differential operator set $(x,\partial_x,\mu)$ and the $q$-differential operator set $(z,\partial_q,\hat\mu)$ (QS-MT correspondence).

Let us consider the correspondence between the QSS bosonic fundamental operator $B_n$ in \eqref{Bn} and the $q$-differential representation \eqref{Ln_diff} of $L_n$:
\beq
B_n=-q^{-1}x^{n+1}\partial_x\quad \leftrightarrow \quad \hat{L}_n=-z^{n+1}\partial_q\,.
\label{B2Lhat}
\eeq
From the formula 
\beq
q\partial_q z=1+q^{-1}z\partial_q\,,\quad 
\eeq
and the bosonic differential part of the $GL_q(1,1)$ quantum plane algebra
\beq
\partial_x x=1+q^{-2}x\partial_x\,,\quad 
\eeq
we recognize that the correspondence \eqref{B2Lhat} implies
\beq
(x,\partial_x) \quad \leftrightarrow \quad (z, q\partial_q) \,. \label{qpl2qdif}
\eeq
This shows that the differential operators in QS correspond to $q$-differential operators, and the $q$-differential operator $\hat{L}_n$ corresponds to the MT operator $\Op{T}{n}{k}$. The differential operators in the $GL_q(1,1)$ quantum space can thus be transformed into the representations of MT operators.

Specifically from \eqref{Tnk20},\eqref{Lpm1} and \eqref{Ln_diff}, $\hat{L}_n$ can be decomposed as follows: 
\begin{align}
&\hat{L}_n=\hat{B}_n+\hat{J}_n \,, \label{LbyBJ}\\
&\hat{B}_n=-\Op{T}{n}{0}=\frac{-z^n}{\qqi} \,,  \label{BTz} \\
&\hat{J}_n=q^{n+2\Delta}\Op{T}{n}{2}=z^n\frac{q^{-2z\partial}}{\qqi}\,. \label{JTz}
\end{align}
The MT commutation relations are
\beq
[\Op{T}{n}{k}, \Op{T}{m}{l}]_{(m-n)}=[\frac{n(l-2)-m(k-2)}{2}]\Op{T}{n+m}{k+l}\,,
\eeq
and \eqref{LbyBJ}, \eqref{BTz} and \eqref{JTz} form the isomorphic to the $CZ_{QS}$ 
relations \eqref{BBBFq} and \eqref{FFU1q}.
The general form of the CZ operator is given by the composition rule \eqref{CZgen}:
\beq
\hat{L}_n^{CZ}=\hat{B}_n - g_n^{CZ}\hat{J}_n \,,  \label{CZgen2}
\eeq
which satisfies the CZ algebra ($CZ_{MT}$), and substituting $g'_n$ instead of $g_n^{CZ}$ gives the $\scz$ algebra.
\eqref{LbyBJ} corresponds to a special case of these ($a=0,b=-1$). Though the relation \eqref{qpl2qdif} represents the QS-MT correspondence, and while $(B_n,J_n)$ and $(\hat{B}_n,\hat{J}_n)$ play similar roles as the constituent elements of the CZ operators, it is clear from \eqref{BTz} and \eqref{JTz} that they do not coincide in the $q\ra1$ limit. Thus, we cannot simply apply this correspondence, and there must be a relationship between $(B_n,J_n)$ and $(\hat{B}_n,\hat{J}_n)$ that is transformed by a mixing matrix. As a result, the algebra in the $q\ra1$ limit may differ from the one expected from the QSS viewpoint. Therefore, for the sake of simplicity, we consider a special mixing situation suggested by \cite{super3,HHT} where the supercharge becomes $q$-anticommutative in the $q\ra1$ limit, while aiming for the super CZ in QSS for $q\not=1$.

Before that, let's notice the corresponding MT counterpart $\hat{\mu}$ of the QS scaling operator $\mu$ in Eq.\eqref{mubyL0}. From \eqref{Ln_diff} and \eqref{qpl2qdif}, 
we have
\beq
\hat{\mu}=1+(\qqi)\hat{L}_0=q(\partial_qz-z\partial_q)=q^{-2z\partial}\,.  \label{muhat}
\eeq
This gives the transmutation rule between $\hat{J}_n$ and $\hat{B}_n$ as
\beq
\hat{J}_n=\frac{z^n\hat{\mu}}{\qqi}=-\hat{B}_n\hat{\mu}\,,  \label{JBmu}
\eeq
and exhibits the scaling rule
\beq
\hat{\mu}\hat{X}_n\hat{\mu}^{-1}=q^{-2n}\hat{X}_n\,,\quad 
\hat{X}_n=\hat{L}_n,\hat{B}_n,\hat{J}_n \,.
\eeq

\subsection{verification}\label{sec:proof}
\indent

To determine the MT version of supercharge $\hat{\mathcal{G}}_r$,
we will parameterize the general form as follows and
determine the parameters from algebraic consistency:
\beq
\mathcal{\hat{G}}_r=\hat{\mu}^{-\nu} \tilde{B}_r \otimes\sigma_1+c
\hat{\mu}^{\nu} \tilde{J}_r \otimes\sigma_2 \,, \label{setG}
\eeq
\beq
\tilde{B}_r= q^\gamma \hat{B}_{r-\frac{1}{2}}\,, \quad\,
\tilde{J}_r=-(q^\alpha \hat{J}_{r+\frac{1}{2}} + q^\beta \hat{B}_{r+\frac{1}{2}})\,.
\label{BJtilde}
\eeq
In the cross terms of the $q$-anticommutator,
we set inverse powers of $\hat{\mu}$ in the first and second terms of \eqref{setG} to cancel each other out. Anticipating two types of products 
$\hat{J}_{r+\frac{1}{2}}\hat{B}_{s-\frac{1}{2}}\sim\hat{J}_{r+s}$ and
$\hat{B}_{r+\frac{1}{2}}\hat{B}_{s-\frac{1}{2}}\sim\hat{B}_{r+s}$
that will form the basis of $L_{r+s}^{CZ}$,
we set $\tilde{J}_r$ as a linear combination of $\hat{J}_{r+\frac{1}{2}}$ and $\hat{B}_{r+\frac{1}{2}}$.
Since $(B_n,J_n)$ and $(\hat{B}_n,\hat{J}_n)$ do not coincide as $q\ra1$, we assume an appropriate mixing relation as mentioned above. Identifying $B_n$'s counterpart with 
$\hat{B}_n$, we set the second term of \eqref{setG} to be proportional to $c$ so that
the supercharge becomes ($q$-)anticommutative as $q\ra1$.

As further assumptions, we consider $\nu$ to be independent of $r$, and $\alpha$, $\beta$, and $\gamma$ to be dependent on $r$:
\beq
\alpha=\alpha(r)\,,\quad\beta=\beta(r)\,,\quad \gamma=\gamma(r)\,,
\eeq
and in the notation for $\mathcal{\hat{G}}s$, we use $\alpha'=\alpha(s)$ and so on.
Using the fusion rule \eqref{Trans}, the calculations can be performed as follows:
\begin{align}
&\hat{B}_{r+\frac{1}{2}}\hat{B}_{s-\frac{1}{2}} = \Op{T}{r+\frac{1}{2}}{0}\Op{T}{s-\frac{1}{2}}{0}
=\frac{1}{\qqi}\Op{T}{r+s}{0}=\frac{-1}{\qqi}\hat{B}_{r+s} \,,\\
&\hat{J}_{r+\frac{1}{2}}\hat{B}_{s-\frac{1}{2}} = -q^{r+\frac{1}{2}+2\Delta}\Op{T}{r+\frac{1}{2}}{2}\Op{T}{s-\frac{1}{2}}{0}=\frac{-q^{r-s+1+2\Delta}}{\qqi}\Op{T}{r+s}{2}=\frac{-q^{1-2s}}{\qqi}\hat{J}_{r+s}\,.
\end{align}
Then, we can obtain the form of
\beq
\{\mathcal{\hat{G}}_r,\mathcal{\hat{G}}_s  \}_{(\frac{s-r}{2})}=c\begin{pmatrix}
\mathcal{L}_G^+ & 0  \\
0 & \mathcal{L}_G^-  \\
\end{pmatrix} \,,   \label{GGeq}
\eeq
where $\mathcal{L}G^\pm$ take the form:
\begin{align}
&\mathcal{L}_G^+=\frac{q^{A_+}+q^{B_+}}{\qqi}q^{-2(r+s)}\hat{J}_{r+s}
          +\frac{q^{C_+}+q^{D_+}}{\qqi}\hat{B}_{r+s}\,,\\
&\mathcal{L}_G^-=\frac{q^{A_-}+q^{B_-}}{\qqi}q^{-1-r-s}\hat{J}_{r+s}
         +\frac{q^{C_-}+q^{D_-}}{\qqi}\hat{B}_{r+s}\,.
\end{align}
Factorization occurs only when
\beq
(A_+,B_+)=(C_+, D_+)\,,\quad\mbox{and}\quad (A_-,B_-)=(D_-,C_-)
\eeq
in which case we have:
\begin{align}
&\mathcal{L}_G^+=A^+(r,s) (q^{-2(r+s)}\hat{J}_{r+s} +\hat{B}_{r+s}) \,,\\
&\mathcal{L}_G^-=A^-(r,s) (q^{-1-r-s}\hat{J}_{r+s} +\hat{B}_{r+s})\,,\\
&A^\pm(r,s)=\frac{q^{A_\pm}+q^{B_\pm}}{\qqi}\,.
\end{align}
Solving \eqref{GGeq} for $\alpha$, $\beta$, and $\gamma$, with $e_1$ and $e_2$ as undetermined constants and $\nu$ as a free parameter, we obtain:
\beq
\alpha(r)=-r-\frac{1}{2}+e_1\,,\quad \beta(r)=r+\frac{1}{2}+e_1\,,\quad \gamma(r)=(1-2\nu)r+e_1+e_2\,, \label{abc_def}
\eeq
and by setting
\beq
e_1=0\,,\quad e_2=\nu+\frac{1}{2} \label{e1e2_def}
\eeq
we get:
\begin{align}
&\mathcal{L}_G^+=q^{1+r+s}[\frac{r-s}{2}]_+ (q^{-2(r+s)}\hat{J}_{r+s} +\hat{B}_{r+s}) q^{-2\nu(r+s)}\,,\\
&\mathcal{L}_G^-=q^{1+r+s}[\frac{r-s}{2}]_+ (q^{-1-r-s}\hat{J}_{r+s} +\hat{B}_{r+s})\,,\\
&[\frac{r-s}{2}]_+=\frac{q^{\frac{r-s}{2}}+q^{\frac{s-r}{2}}}{\qqi}\,.
\end{align}
The $\mathcal{L}_G^{\pm}$ are proportional to the $CZ$ operator \eqref{CZgen} and the $\scz$ operator \eqref{CZTgen}, respectively
\beq
\mathcal{\hat{L}}_n^+=\hat{B}_n+q^{-2n}\hat{J}_n\,,\quad
\mathcal{\hat{L}}_n^-=\hat{B}_n+q^{-1-n}\hat{J}_n\,. \label{supLpm}
\eeq

In summary, we derive the matrix algebra
\beq
\{\mathcal{\hat{G}}_r,\mathcal{\hat{G}}_s  \}_{(\frac{s-r}{2})}=c[\frac{r-s}{2}]_+q^{1+r+s}
\mathcal{\hat{L}}_{r+s}\,, \label{newGGL}
\eeq
\beq
\mathcal{\hat{L}}_n=\begin{pmatrix}
q^{-2n\nu}\mathcal{\hat{L}}_n^+ & 0  \\
0 & \mathcal{\hat{L}}_n^-  \\
\end{pmatrix} \,, \label{supLmat}
\eeq
where the $\mathcal{\hat{L}}_n^\pm$ satisfy
\beq
[\mathcal{\hat{L}}_n^+,\mathcal{\hat{L}}_m^+]_{(m-n)}=[n-m]\mathcal{\hat{L}}_{n+m}^+\,,\quad
[\mathcal{\hat{L}}_n^-,\mathcal{\hat{L}}_m^-]_{(m-n)}=[n-m]\mathcal{\hat{L}}_{n+m}^- +a_{n,m}\hat{J}_{n+m}\,. \label{CZCZ'}
\eeq
Hence the matrix $\mathcal{\hat{L}}_n$ satisfies the super $\scz$ algebra (with $a'=q^{-1}$ from \eqref{supLpm}), and the following matrix relations hold:
\beq
[\mathcal{\hat{L}}_n,\mathcal{\hat{L}}_m]_{(m-n)}=[n-m]\mathcal{\hat{L}}_{n+m}
 +\frac{1}{c}a_{n,m}\mathcal{\hat{J}}_{n+m} \,,\label{LLtwist}
\eeq
\beq
[\mathcal{\hat{L}}_n,\mathcal{\hat{G}}_r]_{(r-\frac{n}{2})}=q^{-n}[\frac{n}{2}-r]\mathcal{\hat{G}}_{r+s}\,. \label{N1LG}
\eeq
Furthermore, the matrix representation of $\hat{J}_n$
\beq
\mathcal{\hat{J}}_n=c\hat{J}_{n}\otimes\sigma_1\sigma_2 \label{Jmat}
\eeq
satisfies the same commutation relations as \eqref{FFU1q}:
\begin{align}
& [\mathcal{\hat{L}}_n,\mathcal{\hat{J}}_m]_{(m-n)}=-q^{-n}[m] \mathcal{\hat{J}}_{n+m}\,,  \label{calLJJJ}  \\
& [\mathcal{\hat{J}}_n,\mathcal{\hat{J}}_m]_{(m-n)}=0\,.  
\end{align}
This corresponds to the super $\scz$ representation in the MT-SM framework. Specifically, \eqref{LLtwist} and \eqref{N1LG} match the QSS representation in \eqref{CZtwist} and \eqref{LnGr2}, and \eqref{newGGL} matches \eqref{GrGs2} except for the overall factor of $q^{3/2}$, which can be absorbed by redefining 
$\mathcal{\hat{G}}_r$.

\section{$N=2$ decomposition}\label{sec:N2}
\indent

If we decompose the supercharge as
\beq
\mathcal{\hat{G}}_r=\mathcal{\hat{G}}_r^+ +\mathcal{\hat{G}}_r^-\,,\quad
\mathcal{\hat{G}}_r^+=\hat{\mu}^{-\nu}\tilde{B}_r\otimes\sigma_1\,,\quad
\mathcal{\hat{G}}_r^-=c\hat{\mu}^{\nu}\tilde{J}_r\otimes\sigma_2\,,
\label{Gmat}
\eeq
then for any constant $p_1$, we have
\beq
\{\mathcal{\hat{G}}_r^\pm,\mathcal{\hat{G}}_r^\pm\}_{(p_1)}=0\,.
\eeq
Calculating the remaining anticommutator to determine $p_1$, we get:
\beq
\{\mathcal{\hat{G}}_r^+,\mathcal{\hat{G}}_s^-\}_{(p_1)}=\begin{pmatrix}
\mathscr{L}_G^+ & 0  \\
0 & \mathscr{L}_G^-  \\
\end{pmatrix} \,,  \label{GGmatrix}
\eeq
where
\begin{align}
&\mathscr{L}_G^+=q^{p_1+2+r+s+(\nu-1)(1+2s)}
         q^{-2\nu(r+s)}\mathcal{\hat{L}}_{r+s}^+\,,\\
&\mathscr{L}_G^-=q^{-p_1+2+r+s+(\nu-1)(1-2r)}
          (\hat{B}_{r+s}+q^{-1-2s}\hat{J}_{r+s})\,.
\end{align}
To factor out the matrix $\mathcal{\hat{L}}_{r+s}$ from the right-hand side of \eqref{GGmatrix}, we need to choose $p_1$ such that the overall factors in $\mathscr{L}_G^\pm$ match:
\beq
p_1=-(r+s)(\nu-1)\,. \label{p1def}
\eeq
Then $\mathscr{L}_G^\pm$ reducing to the form containing diagonal elements of $\mathcal{\hat{L}}_{r+s}$:
\begin{align}
&\mathscr{L}_G^+=q^{2+r+s+(\nu-1)(1+s-r)}
          q^{-2\nu(r+s)}\mathcal{\hat{L}}_{r+s}^+\,,\\
&\mathscr{L}_G^-=q^{2+r+s+(\nu-1)(1+s-r)}\mathcal{\hat{L}}_{r+s}^-
+cq^{(\nu-1)(1+s-r)}q^{\frac{2+r-s}{2}}[\frac{r-s}{2}]\hat{J}_{r+s}\,,
\end{align}
the final expression of \eqref{GGmatrix} is summarized in a closed form using $\mathcal{\hat{L}}_n$ and $\mathcal{\hat{J}}_n$:
\beq
\{\mathcal{\hat{G}}_r^+,\mathcal{\hat{G}}_s^-\}_{(p_1)}=
q^{2+r+s}q^{(\nu-1)(1+s-r)} \mathcal{\hat{L}}_{r+s}
+q^{(\nu-1)(1+s-r)}q^{\frac{2+r-s}{2}}[\frac{r-s}{2}]\mathcal{\hat{J}}_{r+s}\,,
\label{hatGG}
\eeq
\beq
[\mathcal{\hat{L}}_n,\mathcal{\hat{G}}_r^\pm]_{(r-\frac{n}{2})}=q^{-n}[\frac{n}{2}-r]\mathcal{\hat{G}}_{r+s}^\pm\,. \label{hatLG}
\eeq
Comparing to the QSS case, \eqref{hatLG} coincides with \eqref{N2LG}, and \eqref{hatGG} matches \eqref{N2GG} when setting $\nu=1$ and multiplying the right-hand side by $q^{\frac{1}{2}}$. (The phase differences can be absorbed by 
redefinition of $\mathcal{\hat{G}}_r^\pm$).

Finally, we derive the commutation relations between $\mathcal{\hat{J}}_n$ and $\mathcal{\hat{G}}_r^\pm$:
\begin{align}
&[\mathcal{\hat{J}}_n,\mathcal{\hat{G}}_r^+]_{(p_2)}=
      c q^{p_2-2n\nu}\hat{\mu}^{-\nu}\hat{J}_n\tilde{B}_r\otimes\sigma_1\,,\label{JG+}\\
&[\mathcal{\hat{J}}_n,\mathcal{\hat{G}}_r^-]_{(p_2)}=
       -c^2q^{-p_2}\hat{\mu}^{\nu}\tilde{J}_r\hat{J}_n\otimes\sigma_2\,,\label{JG-}
\end{align}
where the phase factor $p_2$ will be determined from the symmetry of the structure constants.

Using the relations \eqref{JBmu} and \eqref{Trans}, we have in a form with $\hat{\mu}$ factored out:
\beq
\hat{J}_{r+\frac{1}{2}}\hat{J}_n=\frac{q^{1+2r}}{\qqi} \hat{\mu}\hat{J}_{n+r+\frac{1}{2}}\,,
\quad\quad
\hat{B}_{r+\frac{1}{2}}\hat{J}_n=\frac{q^{1+2(n+r)}}{\qqi} \hat{\mu}\hat{B}_{n+r+\frac{1}{2}}\,,
\eeq
and using these relations we get
\beq
\tilde{J}_r\hat{J}_n=\frac{q^{n+2r+1}}{\qqi} \hat{\mu}\tilde{J}_{n+r}\,, \quad\quad
\hat{J}_n\tilde{B}_r=\frac{q^{n(1+2\nu)}}{\qqi} \hat{\mu}\tilde{B}_{n+r}\,.
\eeq
Substituting these into \eqref{JG+} and \eqref{JG-}, we obtain
\begin{align}
&[\mathcal{\hat{J}}_n,\mathcal{\hat{G}}_r^+]_{(p_2)}=
       q^{p_2+n}\hat{\mu}\mathcal{\hat{G}}_{n+r}^+\,,\\
&[\mathcal{\hat{J}}_n,\mathcal{\hat{G}}_r^-]_{(p_2)}=
       -q^{-p_2+n+2r+1}\hat{\mu}\mathcal{\hat{G}}_{n+r}^-\,.
\end{align}
Considering the correspondence to \eqref{N2JG} in the QSS case, and redefining 
$\mathcal{\hat{J}}_n^{'}=q^{w}\mathcal{\hat{J}}_n$ from isomorphism, 
we can choose $p_2=r-\frac{n}{2}$ to get:
\beq
[\mathcal{\hat{J}}_n^{'},\mathcal{\hat{G}}_r^\pm]_{(r-\frac{n}{2})}=\pm
       q^{n+r+\frac{1}{2}\mp\frac{n+1}{2}+w}\hat{\mu}\mathcal{\hat{G}}_{n+r}^\pm\,, 
 \label{JGpm}
\eeq
When we set in \eqref{N2JG} requiring consistency with \eqref{JGpm}, 
\beq
\alpha=-\beta=r-\frac{n}{2}-w' \,,
\eeq
we find the unique solution $w=w'=1 \,.$
At this point, \eqref{JGpm} coincides with the QSS $q$-commutation relation in \eqref{N2JG}, describing
\beq
[J_n, G_r^\pm]_{(r-\frac{n}{2})}=\pm
       q^{n+r+\frac{3}{2}\mp\frac{n+1}{2}} \lambda G_{n+r}^\pm\,.
\eeq

\section{Conclusions}
From the above, we have been able to reproduce the $GL_q(1,1)$ super $\scz$ algebra \eqref{CZtwist}\eqref{LnGr2}\eqref{GrGs2} using the MSB representation.
Moreover, in the case of $\nu=1$, we reproduce the $GL_q(1,1)$ $N=2$ super $\scz$ algebra \eqref{N2GG}\eqref{N2LG}\eqref{N2JG}.

We summarize and organize the final results, including the redefinition of the notation.
The representation \eqref{Gmat} of the supercharge obtained as the MSB representation and the superalgebra are as follows:
\[
\mathcal{\hat{G}}_r=\mathcal{\hat{G}}_r^+ +\mathcal{\hat{G}}_r^-\,,\quad
\mathcal{\hat{G}}_r^+=\hat{\mu}^{-\nu}\tilde{B}_r\otimes\sigma_1\,,\quad
\mathcal{\hat{G}}_r^-=c\hat{\mu}^{\nu}\tilde{J}_r\otimes\sigma_2\,,
\]
where $\tilde{B}_r$ and $\tilde{J}_r$ are given by \eqref{BJtilde} 
with \eqref{abc_def} and \eqref{e1e2_def}:
\beq
\tilde{B}_r= q^{(1-2\nu)r+\nu+\frac{1}{2}} \hat{B}_{r-\frac{1}{2}}\,,\quad
\tilde{J}_r=-q^{-r-\frac{1}{2}} \hat{J}_{r+\frac{1}{2}} - q^{r+\frac{1}{2}} \hat{B}_{r+\frac{1}{2}}\,.
\eeq

The MSB representation of $CZ$ generators given by \eqref{supLmat} 
and $U(1)$ current $\mathcal{\hat{J}}_n$ given by \eqref{Jmat} with remembering the redefinition:
\beq
\mathcal{\hat{L}}_n=q^{-2n\nu}\mathcal{\hat{L}}_n^+ \otimes\sigma_2\sigma_1
+\mathcal{\hat{L}}_n^- \otimes\sigma_1\sigma_2
\eeq
\beq
\mathcal{\hat{J}}_n=qc\hat{J}_{n}\otimes\sigma_1\sigma_2
\eeq
where $CZ$ and $\scz$ generators $\mathcal{\hat{L}}_n^\pm$ are given by \eqref{supLpm}.
These satisfy the following $N=1$ super $\scz$ algebra as shown in \eqref{newGGL},\eqref{LLtwist},\eqref{N1LG} and \eqref{calLJJJ}:
\beq
[\mathcal{\hat{L}}_n,\mathcal{\hat{L}}_m]_{(m-n)}=[n-m]\mathcal{\hat{L}}_{n+m}
 +cq^{-1}\, \alpha_{n,m}\mathcal{\hat{J}}_{n+m} \,, \label{superLL}
\eeq
\beq
\{\mathcal{\hat{G}}_r,\mathcal{\hat{G}}_s  \}_{(\frac{s-r}{2})}=
(q^{\frac{r-s}{2}}+q^{\frac{s-r}{2}}) q^{1+r+s}\mathcal{\hat{L}}_{r+s}\,, 
\eeq
\beq
[\mathcal{\hat{L}}_n,\mathcal{\hat{G}}_r]_{(r-\frac{n}{2})}=q^{-n}[\frac{n}{2}-r]\mathcal{\hat{G}}_{r+s}\,, 
\eeq
\beq
 [\mathcal{\hat{L}}_n,\mathcal{\hat{J}}_m]_{(m-n)}=-q^{-n}[m] \mathcal{\hat{J}}_{n+m}\,,  \quad
 [\mathcal{\hat{J}}_n,\mathcal{\hat{J}}_m]_{(m-n)}=0\,,
\eeq
where the diagonal elements $\mathcal{\hat{L}}_n^\pm$ satisfy \eqref{CZCZ'}, 
representing the $CZ$ and $\scz$ algebras respectively:
\beq
[\mathcal{\hat{L}}_n^+,\mathcal{\hat{L}}_m^+]_{(m-n)}=[n-m]\mathcal{\hat{L}}_{n+m}^+\,,\quad
[\mathcal{\hat{L}}_n^-,\mathcal{\hat{L}}_m^-]_{(m-n)}=[n-m]\mathcal{\hat{L}}_{n+m}^- +c^2\,\alpha_{n,m} \hat{J}_{n+m}\,,   \label{sumLL}
\eeq
and $\alpha_{n,m}$ is redefined as $c^{-2}a_{n,m}$ \eqref{a_nm} (with $a'=q^{-1}$):
\beq
\alpha_{n,m}= q^{-1} q^{-n-m}
[\frac{n-m}{2}][\frac{n}{2}][\frac{m}{2}] \,.
\eeq

The $N=2$ superalgebra \eqref{hatGG},\eqref{hatLG},\eqref{JGpm}:
\beq
\{\mathcal{\hat{G}}_r^+,\mathcal{\hat{G}}_s^-\}_{(p_1)}=
q^{2+r+s}q^{(\nu-1)(1+s-r)} \mathcal{\hat{L}}_{r+s}
+q^{(\nu-1)(1+s-r)}q^{\frac{r-s}{2}}[\frac{r-s}{2}]\mathcal{\hat{J}}_{r+s}\,,
\label{GrGsp1}
\eeq
\beq
[\mathcal{\hat{L}}_n,\mathcal{\hat{G}}_r^\pm]_{(r-\frac{n}{2})}=q^{-n}[\frac{n}{2}-r]\mathcal{\hat{G}}_{r+s}^\pm\,. 
\eeq
\beq
[\mathcal{\hat{J}}_n,\mathcal{\hat{G}}_r^\pm]_{(r-\frac{n}{2})}=\pm
       q^{n+r+\frac{3}{2}\mp\frac{n+1}{2}}\hat{\mu}\mathcal{\hat{G}}_{n+r}^\pm\,,
\eeq
where $p_1$ is given by \eqref{p1def}.

Various pieces of knowledge that had been understood fragmentarily are gradually connecting through these lines. The insights gained in this paper are no exception. 
We are now more confident in perceiving the relationships between superalgebras that were once thought entirely unrelated, as well as the approaches and perspectives on their connections to quantum spaces and physical models. Accumulating these algebraic properties serves as a powerful tool, providing clues for further considerations, leading to problem-solving, and driving the pursuit of truth.

\section*{Declaration of generative AI and AI-assisted technologies in the writing process}
During the preparation of this work the author used Claude 3.5 Sonnet in order to improve grammar and enhance language expression. 
After using these services, the author reviewed and edited the content as needed and takes full responsibility for the content of the publication.

\section*{CRediT authorship contribution statement} 
 \textbf{Haru-Tada Sato:} Writing-Original Draft, Conceptualization, Methodology, Investigation, Validation.

\section*{Funding sources}
This research did not receive any specific grant from funding agencies in the public, commercial, or not-for-profit sectors.

\section*{Declaration of competing interest}
The author declares that we have no known competing financial interests or personal relationships that could have appeared to influence the work reported in this paper.


\section*{Data availability}
No data was used for the research described in the article. 



\newcommand{\NP}[1]{{\it Nucl.{}~Phys.} {\bf #1}}
\newcommand{\PL}[1]{{\it Phys.{}~Lett.} {\bf #1}}
\newcommand{\Prep}[1]{{\it Phys.{}~Rep.} {\bf #1}}
\newcommand{\PR}[1]{{\it Phys.{}~Rev.} {\bf #1}}
\newcommand{\PRL}[1]{{\it Phys.{}~Rev.{}~Lett.} {\bf #1}}
\newcommand{\PTP}[1]{{\it Prog.{}~Theor.{}~Phys.} {\bf #1}}
\newcommand{\PTPS}[1]{{\it Prog.{}~Theor.{}~Phys.{}~Suppl.} {\bf #1}}
\newcommand{\MPL}[1]{{\it Mod.{}~Phys.{}~Lett.} {\bf #1}}
\newcommand{\IJMP}[1]{{\it Int.{}~J.{}~Mod.{}~Phys.} {\bf #1}}
\newcommand{\IJTP}[1]{{\it Int.{}~J.{}~Theor.{}~Phys.} {\bf #1}}
\newcommand{\JPA}[1]{{\it J.{}~Phys.} {\bf A}:\ Math.~Gen. {\bf #1}~}
\newcommand{\JHEP}[1]{{\it J.{}~High Energy{}~Phys.} {\bf #1}}
\newcommand{\JMP}[1]{{\it J.{}~Math.{}~Phys.} {\bf #1} }
\newcommand{\CMP}[1]{{\it Commun.{}~Math.{}~Phys.} {\bf #1} }
\newcommand{\LMP}[1]{{\it Lett.{}~Math.{}~Phys.} {\bf #1} }
\newcommand{\doi}[2]{\,\href{#1}{#2}\,}  


\end{document}